\documentclass{svproc}

\usepackage{url}
\usepackage{amsmath}
\usepackage{graphicx}
\usepackage{xcolor}
\usepackage{subcaption}
\usepackage{amsfonts}
\usepackage{multirow}
\usepackage{dsfont} 
\usepackage[strings]{underscore}
\usepackage{hyperref}
\usepackage{cleveref}
\begin{document}
\mainmatter   
\title{Measuring Proximity in Attributed Networks for Community Detection}

\titlerunning{Measuring Proximity in Attributed Networks for Community Detection}  
\author{Rinat Aynulin, Pavel Chebotarev}

\authorrunning{Rinat Aynulin, Pavel Chebotarev}

\author{Rinat Aynulin\inst{1,3} \and Pavel Chebotarev\inst{2}}

\institute{
Moscow Institute of Physics and Technology, 9 Inststitutskii per., Dolgoprudny, Moscow Region, 141700 Russia,
\email{rinat.aynulin@phystech.edu}
\and
Trapeznikov Institute of Control Sciences of the Russian Academy of Sciences, 65 Profsoyuznaya str., Moscow, 117997 Russia, \email{pavel4e@gmail.com}
\and 
Kotel’nikov Institute of Radio-engineering and Electronics (IRE) of the Russian Academy of
Sciences, Mokhovaya 11-7, Moscow, 125009 Russia}
\maketitle              

\begin{abstract}
Proximity measures on graphs have a variety of applications in network analysis, including community detection. Previously they have been mainly studied in the context of networks without attributes. If node attributes are taken into account, however, this can provide more insight into the network structure. In this paper, we extend the definition of some well-studied proximity measures to attributed networks. To account for attributes, several attribute similarity measures are used. Finally, the obtained proximity measures are applied to detect the community structure in some real-world networks using the spectral clustering algorithm\footnote{This is a preprint of the following chapter: Aynulin R., Chebotarev P., Measuring Proximity in Attributed Networks for Community Detection, published in Studies in Computational Intelligence, vol 943, edited by Benito R.M., Cherifi C., Cherifi H., Moro E., Rocha L.M., Sales-Pardo M., 2021, Springer, reproduced with permission of Springer Nature Switzerland AG. The final authenticated version is available online at: \url{http://dx.doi.org/10.1007/978-3-030-65347-7_3}}.

\end{abstract}

\section{Introduction}
Many real-world systems from the fields of social science, economy, biology, chemistry, etc. can be represented as networks or graphs\footnote{Formally, graph is a mathematical representation of a network. However, hereinafter, the terms ``graph'' and ``network'' will be used interchangeably.} \cite{applications}. A network consists of nodes representing objects, connected by edges representing relations between the objects. Nodes can often be divided into groups called clusters or communities. Members of such a cluster are more densely connected to each other than to the nodes outside the cluster.

The task of finding such groups is called clustering or community detection. There have been plenty of algorithms proposed by researchers in the past to address this problem. 

Some of the community detection algorithms require the introduction of distance or proximity measure on the set of graph nodes: a function, which shows, respectively, the distance or proximity (similarity) between a pair of nodes. Only the shortest path distance had been studied for a long time. Nowadays, we have a surprising variety of measures on the set of graph nodes \cite[Chapter 15]{DezaDeza16}. Some of the proximity measures can be defined as kernels on graphs, i.e., symmetric positive semidefinite matrices \cite{cheb-on-kernels}.

Previously, kernels have been applied mainly to analyze networks without attributes. However, in many networks, nodes are associated with attributes that describe them in some way. Thereby, multiple dimensions of information can be available: a structural dimension representing relations between objects, a compositional dimension describing attributes of particular objects, and an affiliation dimension representing the community structure \cite{bothorel2015clustering}. Combining information about relations between nodes and their attributes provides a deeper understanding of the network structure.

Many methods for community detection in attributed networks have been proposed recently. Surveys \cite{bothorel2015clustering,chunaev2020community} describe existing approaches to this problem. We provide some information on this in Section \ref{sec:relwork}. However, kernel-based clustering, as already noted, has not yet been applied to attributed networks.

In this paper, we extend the definition of a number of previously defined proximity measures to the case of networks with node attributes. Several similarity measures on attributes are used for this purpose. Then, we apply the obtained proximity measures to the problem of community detection in several real-world datasets.

According to the results of our experiments, taking both node attributes and node relations into account can improve the efficiency of clustering in comparison with clustering based on attributes only or on structural data only. Also, the most effective attribute similarity measures in our experiments are the Cosine Similarity and Extended Jaccard Similarity.

\section{Related Work}
\label{sec:relwork}
This section is divided into two parts. In the first one, we provide a quick overview of papers where various measures on the set of graph nodes are discussed. Then, we introduce a few studies focused on community detection in attributed networks.

For a long time, only the shortest path distance has been widely used \cite{dijkstra1959note}. \cite[Chapter 15]{DezaDeza16} provides a survey of dozens of measures that have been proposed in various studies in the last decades. Among them there are inspired by physics Resistance (also known as Electric) measure \cite{sharpe1967solution}, logarithmic Walk measure discussed in \cite{cheb-walk}, the Forest measure related to Resistance \cite{cheb-forest-kernel}, and many others.

In \cite{cheb-on-kernels}, the authors analytically study properties of various proximity measures\footnote{Here, we use the term ``proximity measure'' in a broaded sense and, unlike \cite{cheb-on-kernels}, do not require a proximity measure to satisfy the triangle inequality for proximities.} and kernels on graphs, including Walk, Communicability, Heat, PageRank, and several logarithmic measures. Then, these measures are compared in the context of spectral clustering on the stochastic block model. \cite{fouss2012experimental} provides a survey and numerical comparison of nine kernels on graphs in application to link prediction and clustering problems.

In \cite {comparison-felix}, the authors numerically study the efficiency of the Corrected Commute-Time, Free Energy, Logarithmic Forest, Randomized Shortest-Path, Sigmoid Commute-Time, and Shortest-Path measures in experiments with 15 real-world datasets. In \cite{comparison-logarithmic,transformations} it was proposed to improve the efficiency of some existing proximity measures by applying simple mathematical functions like logarithm to them.

Classically, community detection algorithms used either structure information (see, e.g., \cite{girvan2002community}) or information about node attributes (e.g., \cite{jain2010data}). Recently, the idea of detecting communities based both on the structure and attribute data has attracted a lot of attention. Taking into account that it is possible to consider also edge attributes, we will focus on the attributes of nodes.

In \cite{zhou2009graph}, the authors proposed the SA-Clustering algorithm. The idea of the algorithm is the following: first, an attribute node is created for each value of each attribute. An attribute edge is drawn between the ``real'' node and attribute node if the node has the value of the attribute specified in the attribute node. The random walk model then is used to estimate the distance between nodes. Communities are determined using the $k$-medoids method.

The CODICIL method is presented in \cite{ruan2013efficient}. This method adds content edges as a supplement to structure edges. The presence of a content edge between two nodes means the similarity of the node attributes. Then, the graph with content edges is clustered using the Metis and Multi-level Regularized Markov Clustering algorithms.

Reference \cite{neville2003clustering} proposes the method for community detection in attributed networks based on weight modification. For every existing edge, the weight of the edge is assigned to the matching coefficient between the nodes. This coefficient equals to the number of attribute values the nodes have in common. The network with modified edges is clustered using the Karger’s Min-Cut, MajorClust, and Spectral algorithms. In \cite{yang2013community}, the CESNA method is proposed. This method assumes the attributed networks to be generated by a probabilistic model. Communities are detected using maximum-likelihood estimation on this model. 

For a more detailed review of recently proposed methods for community detection in attributed networks, see \cite{bothorel2015clustering,chunaev2020community}.

\section{Background and Preliminaries}
\subsection{Definitions}
Let $G = (V, E, F)$ be an undirected weighted attributed graph with the set of nodes $V$  ($|V|=n$), the set of edges $E$ ($|E|=m$), and the tuple of attribute (or feature) vectors $F$. Each of the $n$ nodes is associated with $d$ attributes, so $F = (\mathbf{f}_1, ..., \mathbf{f}_n)$, where $\mathbf{f}_i \in {\mathbb R}^d$. In the experiments, we will consider networks with binary attributes.

The \textit{adjacency matrix} $A$ of the graph is a square matrix with elements $a_{ij}$ equal to the weight of edge $(i, j)$ if node $i$ is connected to node $j$ and equal to zero otherwise. In some applications, each edge can also be associated with a positive value $c_{ij}$, which is the cost of following this edge. If cost does not appear naturally, it can be defined as $c_{ij} = \frac{1}{a_{ij}}$. The \textit{cost matrix} $C$ contains costs of all the edges.

The \textit{degree} of a node is the sum of the weights of the edges linked to the node. The diagonal \textit{degree matrix} $D = \mathrm{diag} (A \cdot \textbf{1})$ shows degrees of all the nodes in the graph ($\mathbf{1}=(1,...,1)^T$). Given $A$ and $D$, the \textit{Laplacian matrix} is defined as $L = D - A$, and the \textit{Markov matrix} is $P = D^{-1}A$. 

A \textit{measure} on the set of graph nodes is a function $\kappa$ that characterizes proximity or similarity between the pairs of graph nodes. A \textit{kernel on graph} is a similarity measure that has a Gram matrix (symmetric positive semidefinite matrix) representation $K$. Given $K$, the corresponding distance matrix $\rm{\Delta}$ can be obtained from the equation \begin{equation}
\label{eq:distance-to-kernel-transformation}
    K = - \frac{1}{2} H \rm{\Delta} H,
\end{equation}
where $H = I - \frac{1}{n} \textbf{1} \cdot \textbf{1}^T$.

For more details about graph measures and kernels, we refer to \cite{cheb-on-kernels}.

\subsection{Community Detection Algorithms}

\subsubsection{$k$-means.} 
The $k$-means algorithm \cite{macqueen1967some} is used in this study for community detection based on the attribute information. 

\subsubsection{Spectral.} 
In this paper, we use the variation of the Spectral algorithm presented by Shi and Malik in \cite{shi2000normalized}. The approach is based on applying the $k$-means algorithm to the eigenvectors of the Laplacian matrix of the graph. For a detailed review of the mathematics behind the Spectral algorithm, we refer to the tutorial by Ulrike von Luxburg \cite{spectral-tutorial}.

\subsection{Measures}
\label{sec:measures}

In this study, we consider five measures which have shown a good efficiency in \cite{cheb-on-kernels,comparison-felix}.

\subsubsection{Communicability.} $K^\mathrm{C} = \sum_{n=0}^{\infty} \frac{ \alpha ^ n A ^ n}{n!}= \mathrm{exp}(\alpha A)$, $\alpha > 0$ \cite{comm-distance,comm-distance-2}.
\subsubsection{Heat.} $K^\mathrm{H} = \sum_{n=0}^{\infty} \frac{ \alpha ^ n (-L) ^ n}{n!}= \mathrm{exp}(-\alpha L)$, $\alpha > 0$ \cite{heat-kernel}.
\subsubsection{PageRank.} 
$K^{\mathrm{PR}} = (I - \alpha P) ^ {-1}$, $0 < \alpha < 1$ \cite{pagerank,fouss2012experimental}.
\subsubsection{Free Energy.}
Given $P$, $C$ and the parameter $\alpha$, the matrix $W$ can be defined as $W = \rm{exp}(-\alpha C) \circ P$ (the ``$\circ$'' symbol stands for element-wise multiplication). Then, $Z = (I - W)^{-1}$ and $S = (Z (C \circ W)) \div Z$ (the ``$\div$'' symbol stands for element-wise division). Finally, $\mathrm{\Delta}^{\mathrm{FE}} = \frac{\Phi + \Phi^T}{2}$, where $\Phi = \frac{\mathrm{log}(Z)}{\alpha}$. $K^{\mathrm{FE}}$ can be obtained from $\mathrm{\Delta}^{\mathrm{FE}}$ using transformation \eqref{eq:distance-to-kernel-transformation} \cite{kivimaki2014developments}.

\subsubsection{Sigmoid Corrected Commute-Time.} First, let us define the Corrected Commute-Time (CCT) kernel: $K^{\mathrm{CCT}} = HD^{-\frac{1}{2}}M(I - M)^{-1}MD^{-\frac{1}{2}}H$, where $H = I - \frac{\mathbf{1} \cdot \mathbf{1}^T}{n}$, $M = D^{-\frac{1}{2}} (A - \frac{\mathbf{d} \cdot \mathbf{d}^T}{\mathrm{vol}(G)}) D^{-\frac{1}{2}}$, $\mathbf{d}$ is a vector of elements of the diagonal degree matrix $D$, $\mathrm{vol}(G) = \sum_{ij=1}^n a_{ij}$. Then, the elements of $K^{\mathrm{SCCT}}$ are equal to $K^{\mathrm{SCCT}}_{ij} = \frac{1}{1 + \mathrm{exp}(-\alpha K^{\mathrm{CCT}}_{ij}/\sigma)}$, where $\sigma$ is the standard deviation of the elements of $K^{\mathrm{CCT}}$, $\alpha > 0$ \cite{luxburg2010getting,comparison-felix}.

\subsection{Clustering Quality Evaluation}

To evaluate the community detection performance, we employ the Adjusted Rand Index (ARI) introduced in \cite{ari-hubert}. Some advantages of this quality index are listed in \cite{ari-best}.

ARI is based on the Rand Index (RI) introduced in \cite{rand}. The Rand Index quantifies the level of agreement between two partitions of $n$ elements $X$ and $Y$. Given $a$ as the number of pairs of elements that are in the same clusters in both partitions, and $b$ the number of pairs of elements in different clusters in both partitions, the Rand Index is defined as $\frac{a + b}{\binom{n}{2}}$.

The Adjusted Rand Index is the transformation of the Rand Index such that its expected value is 0 and maximum value is 1: $\rm{ARI} = \frac{\rm{Index} - \rm{ExpectedIndex}}{\rm{MaxIndex} - \rm{ExpectedIndex}}$.

\section{Proximity-based Community Detection in Attributed Networks}
\label{sec:clustering-method}

In order to apply the proximity measures described in Section \ref{sec:measures} to attributed networks, we need a way to embed node attribute information into the adjacency matrix. This can be done by modifying edge weights based on the attributes: \begin{equation}\label{eq:similarity-measure}a^s_{ij} = \beta a_{ij} + (1 - \beta) s_{ij},\end{equation}  where $\beta \in [0, 1]$ and $s_{ij} = s(\mathbf{f}_i, \mathbf{f}_j)$ is an attribute similarity measure calculated for nodes $i$ and $j$. An attribute similarity measure, as the name implies, shows to what extent two nodes are similar by attributes.

By varying the coefficient $\beta$, we can make a trade-off between weighted adjacency and attribute similarity. So, when $\beta = 0$, the attributed adjacency matrix $A^s$ describes only nodes similarity by attributes, while with $\beta = 1$ it coincides with $A$. 

Given $A^s$, we can compute attributed versions of all the other matrices required to define proximity measures. Then, the proximity measures can be calculated and applied for detecting clusters using the Spectral method.

To take node attributes into account, we use various attribute similarity measures. Let $\mathbf{f}_i = (f^1_i, ..., f^d_i)$ and $\mathbf{f}_j = (f^1_j, ..., f^d_j)$ be the attribute vectors of nodes $i$ and $j$, respectively. The attribute similarity measures are defined as following:

\begin{itemize}
    \item Matching Coefficient\footnote{Since equality will be rare for continuous attributes, Matching Coefficient is mainly used for discrete attributes, especially binary ones.} \cite{vsulc2014evaluation}: $\displaystyle{s^{\mathrm{MC}}(\mathbf{f}_i, \mathbf{f}_j) = \frac{\sum_{k=1}^d \mathds{1}(f^k_i = f^k_j)}{d}}$, where $\mathds{1}(x)$ is the indicator function which takes the value of one if the condition $x$ is true and zero otherwise;
    \item Cosine Similarity \cite[Chapter 2]{tan2016introduction}: $\displaystyle{s^{\mathrm{CS}}(\mathbf{f}_i, \mathbf{f}_j) = \frac{\mathbf{f}_i \cdot \mathbf{f}_j}{||\mathbf{f}_i||_2 ||\mathbf{f}_j||_2}}$;
    \item Extended Jaccard Similarity \cite[Chapter 2]{tan2016introduction}: $\displaystyle{s^{\mathrm{JS}}(\mathbf{f}_i, \mathbf{f}_j) = \frac{\mathbf{f}_i \cdot \mathbf{f}_j}{||\mathbf{f}_i||_2^2 + || \mathbf{f}_j||_2^2 - \mathbf{f}_i \cdot \mathbf{f}_j}}$;
    \item Manhattan Similarity \cite{dang2012community}: $\displaystyle{s^{\mathrm{MS}}(\mathbf{f}_i, \mathbf{f}_j) = \frac{1}{1 + ||\mathbf{f}_i - \mathbf{f}_j||_1}}$;
    \item Euclidean Similarity \cite{dang2012community}: $\displaystyle{s^{\mathrm{ES}}(\mathbf{f}_i, \mathbf{f}_j) = \frac{1}{1 + ||\mathbf{f}_i - \mathbf{f}_j||_2}}$.
\end{itemize}

\section{Experiments}
In this section, we compare attribute-aware proximity measures with the plain ones in experiments with several real-world datasets:

\begin{itemize}
    \item WebKB \cite{lu2003link}: a dataset of university web pages. Each web page is classified into one of five classes: course, faculty, student, project, staff. Each node is associated with a binary feature vector ($d = 1703$) describing presence or absence of words from the dictionary. This dataset consists of four unweighted graphs: Washington ($n = 230$, $m = 446$), Wisconsin ($n = 265$, $m = 530$), Cornell ($n = 195$, $m = 304$), and Texas ($n = 187$, $m = 328$).
     
    \item CiteSeer \cite{sen2008collective}: an unweighted citation graph of scientific papers. The dataset contains 3312 nodes and 4732 edges. Each paper in the graph is classified into one of six classes (the topic of the paper) and associated with a binary vector ($d = 3703$) describing the presence of words from the dictionary.
    
     \item Cora \cite{sen2008collective}: an unweighted citation graph of scientific papers with a structure similar to the CiteSeer graph. The number of nodes: $n = 2708$, the number of edges: $m = 5429$, the number of classes: $c = 7$, and the number of words in the dictionary (the length of the feature vector): $d = 1433$.
    
\end{itemize}

These datasets are clustered using multiple methods. First, we apply the $k$-means algorithm, which uses only attribute information and ignores graph structure. Then, each dataset is clustered with the Spectral algorithm and five plain proximity measures that do not use attribute information. Finally, communities are detected using the Spectral algorithm and attribute-aware proximity measures that employ both data dimensions (structure and attributes).

We use balanced versions of attribute similarity measures with $\beta = \frac{1}{2}$ in \eqref{eq:similarity-measure}.

Each of the proximity measures depends on the parameter. So, we search for the optimal parameter in the experiments, and the results include clustering quality for the optimal parameter.

\section{Results}
In this section, we discuss the results of the experiments.

In Table \ref{table:experiments-results}, ARI for all the tested proximity measures and similarity measures on all the datasets is presented. ``No'' column shows the result for plain proximity measures that do not use attribute information. The table also presents ARI for the $k$-means clustering algorithm. The top-performing similarity measure is marked in red for each proximity measure.

\begin{table}[]
\caption{Results of the experiments}
\label{table:experiments-results}
\centering
\begin{tabular}{p{2.5cm}p{1.2cm}p{1.2cm}p{1.2cm}p{1.2cm}p{1.2cm}p{1.2cm}}
\hline
\multirow{2}{*}{Prox. Measure}  & \multicolumn{6}{c}{Similarity Measure} \\ \cline{2-7} 
                         & MC   & CS   & JS   & MS   & ES   & No  \\ \hline
\multicolumn{7}{c}{Washington}                                      \\ \hline
\multicolumn{1}{l}{Communicability}&0.048&\textcolor{red}{0.458}&0.352&0.13&0.24&0.05\\
\multicolumn{1}{l}{Heat}&0.058&\textcolor{red}{0.457}&0.444&0.093&0.055&0.043\\
\multicolumn{1}{l}{PR}&0.269&\textcolor{red}{0.461}&0.39&0.097&0.087&0.037\\
\multicolumn{1}{l}{FE}&0.291&\textcolor{red}{0.461}&0.322&0.129&0.243&0.009\\
\multicolumn{1}{l}{SCCT}&0.307&\textcolor{red}{0.397}&0.362&0.132&0.28&0.048\\
\multicolumn{1}{l}{$k$-means}   & \multicolumn{6}{c}{0.095} \\ \hline

\multicolumn{7}{c}{Wisconsin} \\ \hline
\multicolumn{1}{l}{Communicability}&0.089&\textcolor{red}{0.459}&0.41&0.104&0.064&0.02\\
\multicolumn{1}{l}{Heat}&0.075&\textcolor{red}{0.472}&0.416&0.111&0.092&0.082\\
\multicolumn{1}{l}{PR}&0.126&\textcolor{red}{0.471}&0.36&0.13&0.067&0.045\\
\multicolumn{1}{l}{FE}&0.081&\textcolor{red}{0.441}&0.398&0.066&0.066&0.064\\
\multicolumn{1}{l}{SCCT}&0.056&0.354&\textcolor{red}{0.383}&0.103&0.089&0.045\\
\multicolumn{1}{l}{$k$-means}   & \multicolumn{6}{c}{0.364} \\ \hline

\multicolumn{7}{c}{Cornell} \\ \hline
\multicolumn{1}{l}{Communicability}&0.012&\textcolor{red}{0.2}&0.107&0.034&0.058&0.035\\
\multicolumn{1}{l}{Heat}&0.064&\textcolor{red}{0.181}&0.109&0.069&0.046&0.072\\
\multicolumn{1}{l}{PR}&0.047&\textcolor{red}{0.118}&0.088&0.011&-0.025&0.063\\
\multicolumn{1}{l}{FE}&0.053&0.308&\textcolor{red}{0.309}&0.046&0.058&-0.013\\
\multicolumn{1}{l}{SCCT}&0.076&\textcolor{red}{0.193}&0.112&0.057&0.055&0.027\\
\multicolumn{1}{l}{$k$-means}   & \multicolumn{6}{c}{0.066} \\ \hline

\multicolumn{7}{c}{Texas} \\ \hline
\multicolumn{1}{l}{Communicability}&0.118&\textcolor{red}{0.288}&0.281&0.177&0.23&0.008\\
\multicolumn{1}{l}{Heat}&0.137&0.212&\textcolor{red}{0.289}&0.076&0.156&-0.013\\
\multicolumn{1}{l}{PR}&0.221&0.174&\textcolor{red}{0.287}&0.073&0.133&0.0\\
\multicolumn{1}{l}{FE}&0.23&\textcolor{red}{0.342}&0.233&0.23&0.234&0.041\\
\multicolumn{1}{l}{SCCT}&0.274&\textcolor{red}{0.344}&0.258&0.21&0.253&0.15\\
\multicolumn{1}{l}{$k$-means}   & \multicolumn{6}{c}{\textcolor{red}{0.409}} \\ \hline

\multicolumn{7}{c}{CiteSeer} \\ \hline
\multicolumn{1}{l}{Communicability}&0.0&0.24&\textcolor{red}{0.282}&0.001&0.001&0.113\\
\multicolumn{1}{l}{Heat}&0.001&0.258&\textcolor{red}{0.276}&0.005&0.004&0.112\\
\multicolumn{1}{l}{PR}&0.003&0.242&\textcolor{red}{0.266}&-0.0&0.001&0.109\\
\multicolumn{1}{l}{FE}&0.0&\textcolor{red}{0.41}&0.41&0.162&0.186&-0.001\\
\multicolumn{1}{l}{SCCT}&0.001&0.252&\textcolor{red}{0.31}&0.051&0.183&0.018\\
\multicolumn{1}{l}{$k$-means}   & \multicolumn{6}{c}{0.1} \\ \hline

\multicolumn{7}{c}{Cora} \\ \hline
\multicolumn{1}{l}{Communicability}&0.0&0.107&\textcolor{red}{0.119}&0.09&0.027&0.002\\
\multicolumn{1}{l}{Heat}&0.071&\textcolor{red}{0.083}&0.061&0.076&0.024&0.002\\
\multicolumn{1}{l}{PR}&0.032&\textcolor{red}{0.138}&0.135&0.068&0.029&0.005\\
\multicolumn{1}{l}{FE}&0.023&\textcolor{red}{0.408}&0.404&0.301&0.236&-0.001\\
\multicolumn{1}{l}{SCCT}&0.025&0.156&\textcolor{red}{0.189}&0.127&0.184&0.002\\
\multicolumn{1}{l}{$k$-means}   & \multicolumn{6}{c}{0.07} \\ \hline
\end{tabular}
\end{table}

As we can see, taking attributes into account improves community detection quality for all the proximity measures. Attribute-aware proximity measures outperform $k$-means for all the datasets except Texas. Therefore, we can conclude that in most cases, the proximity measures based on structure and attribute information perform better than both the plain proximity measures which use only structure information and the $k$-means clustering method which uses only attribute information. 

\begin{figure}[htp]
\centering

\includegraphics[width=.32\textwidth]{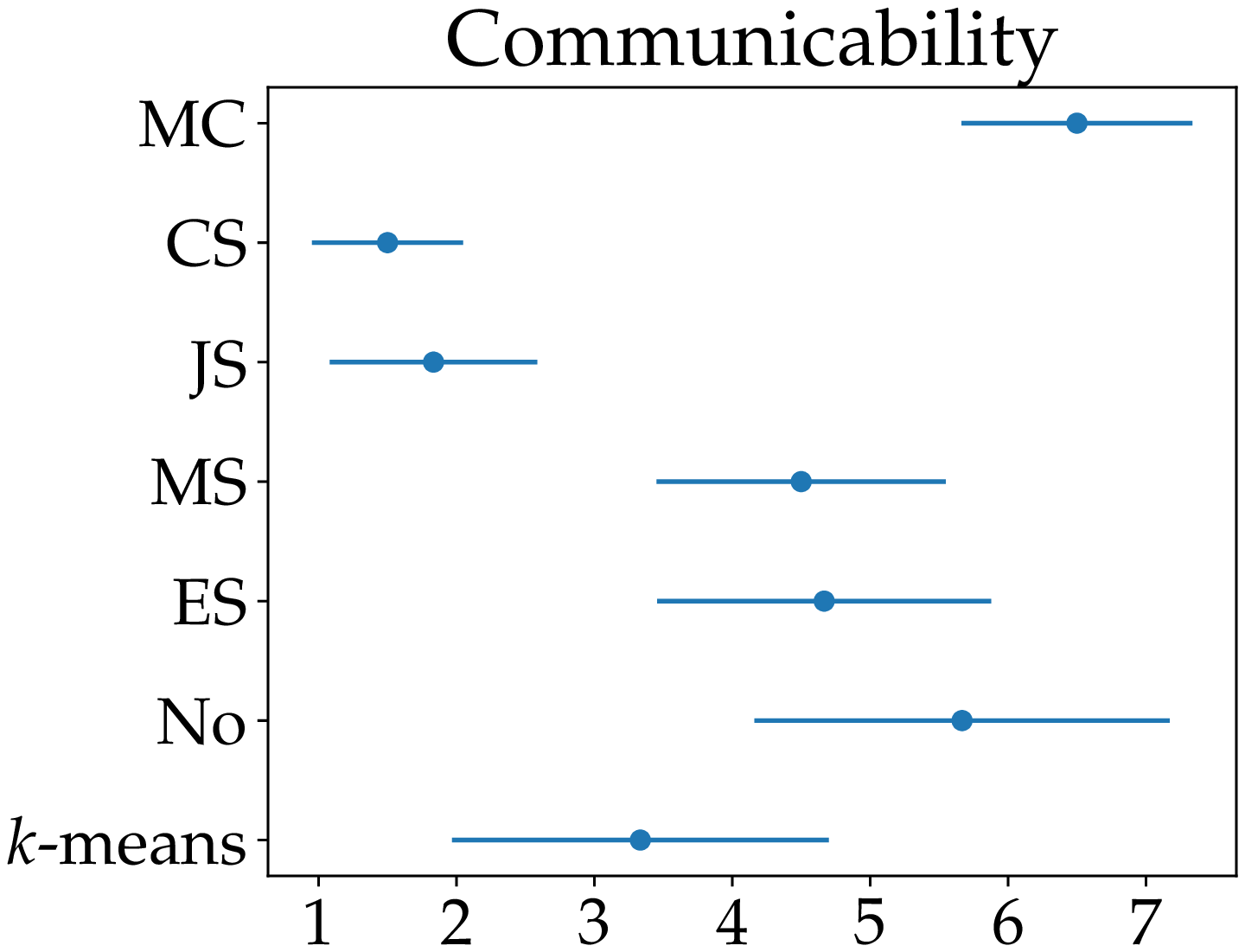}
\includegraphics[width=.32\textwidth]{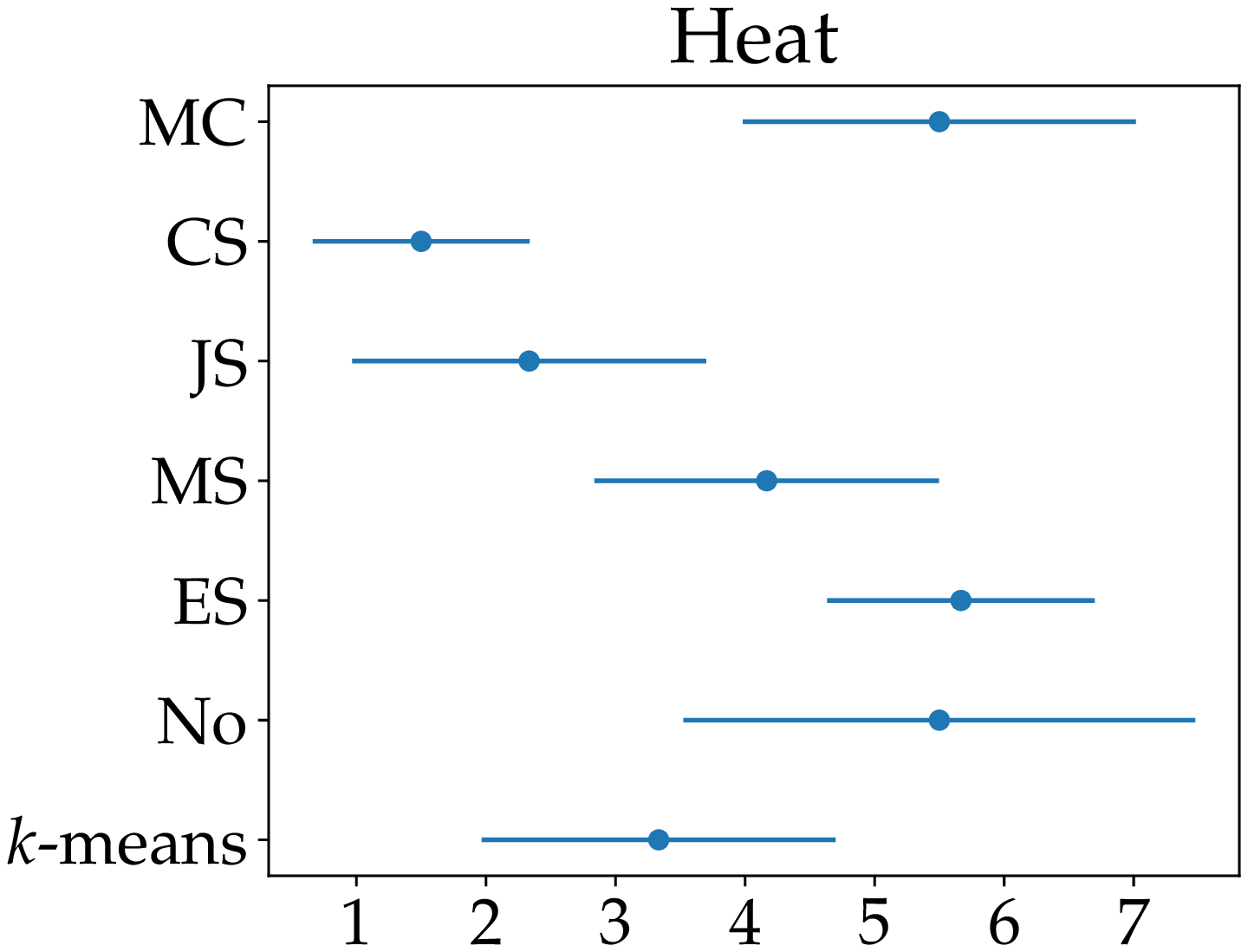}
\includegraphics[width=.32\textwidth]{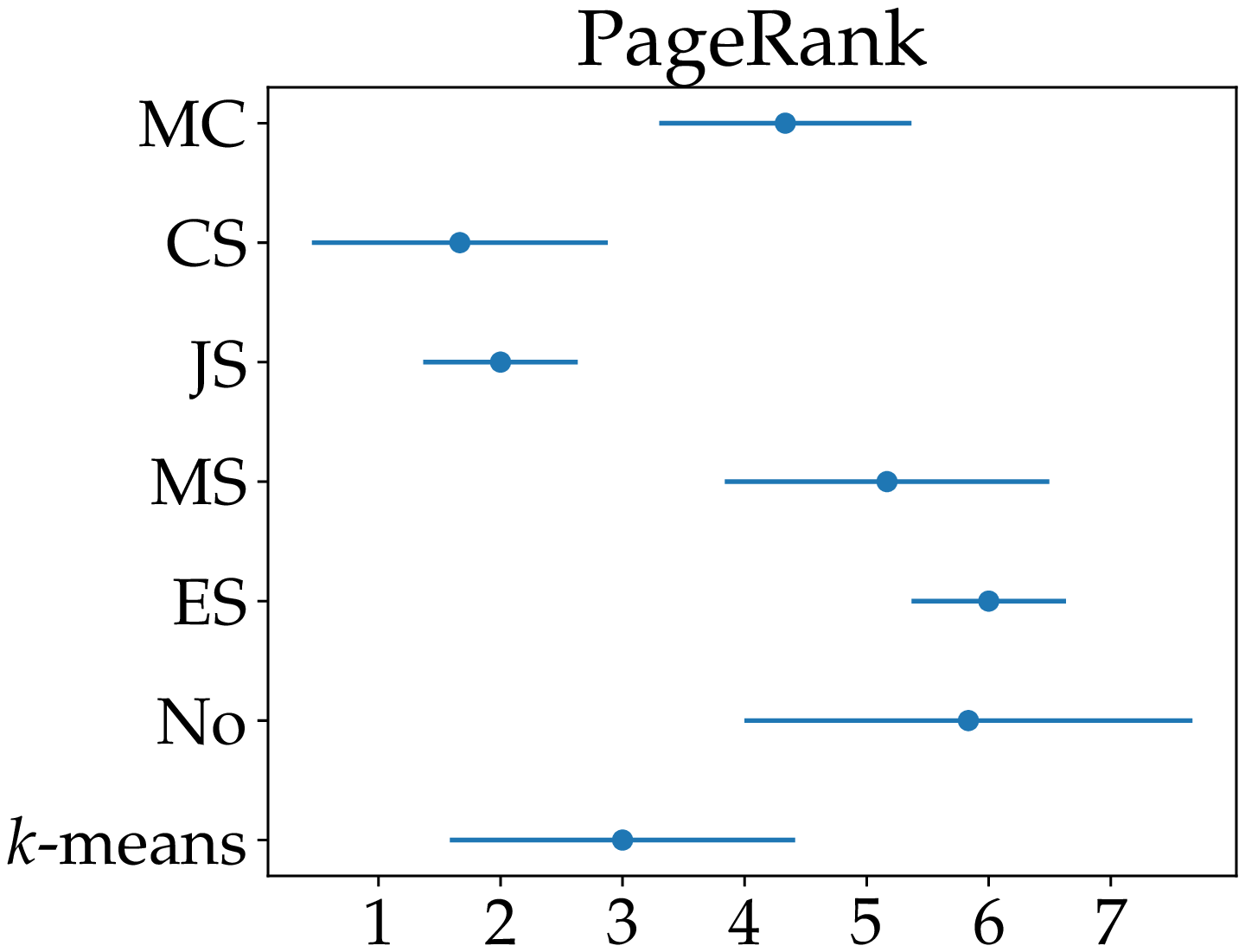}

\medskip

\includegraphics[width=.32\textwidth]{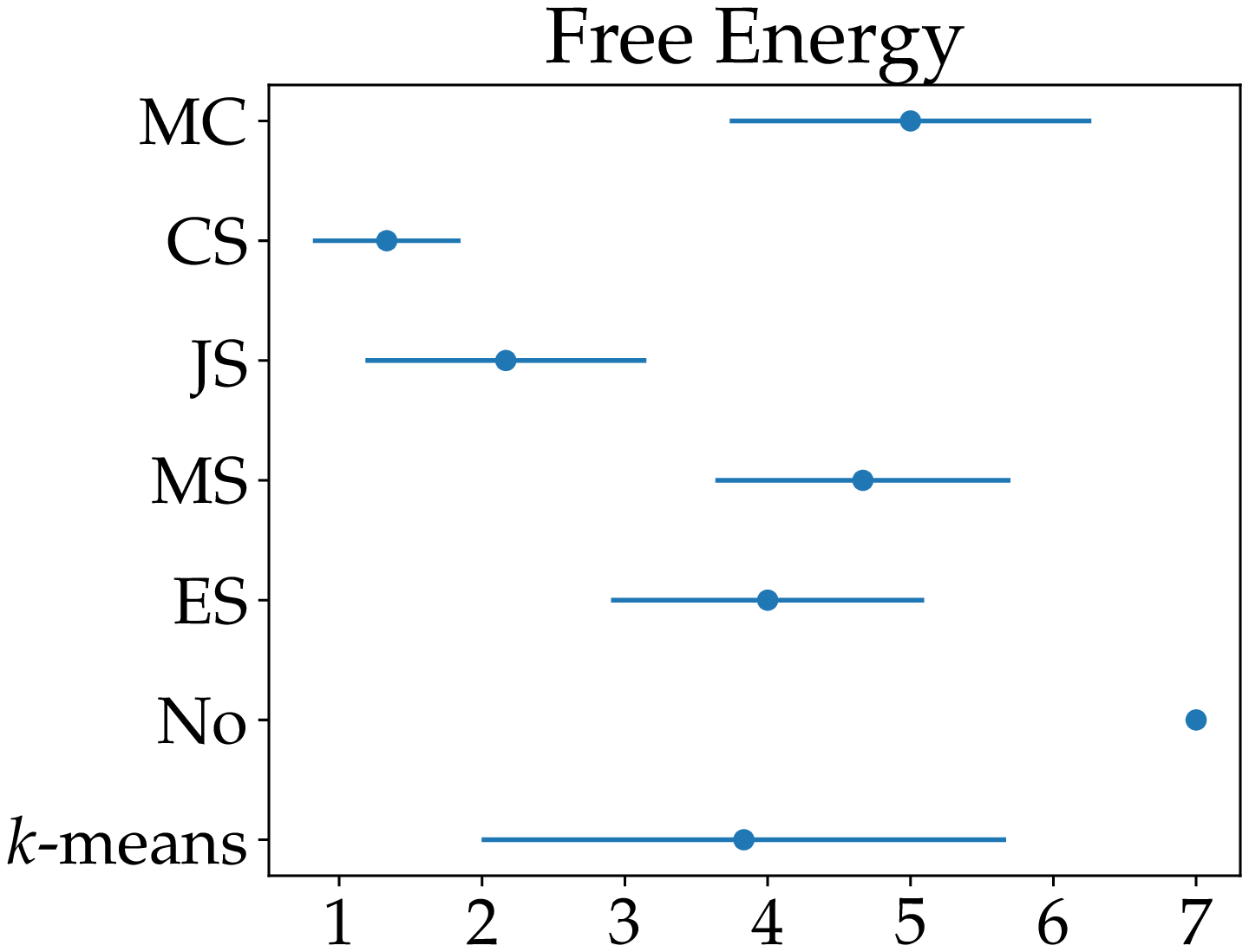}
\includegraphics[width=.32\textwidth]{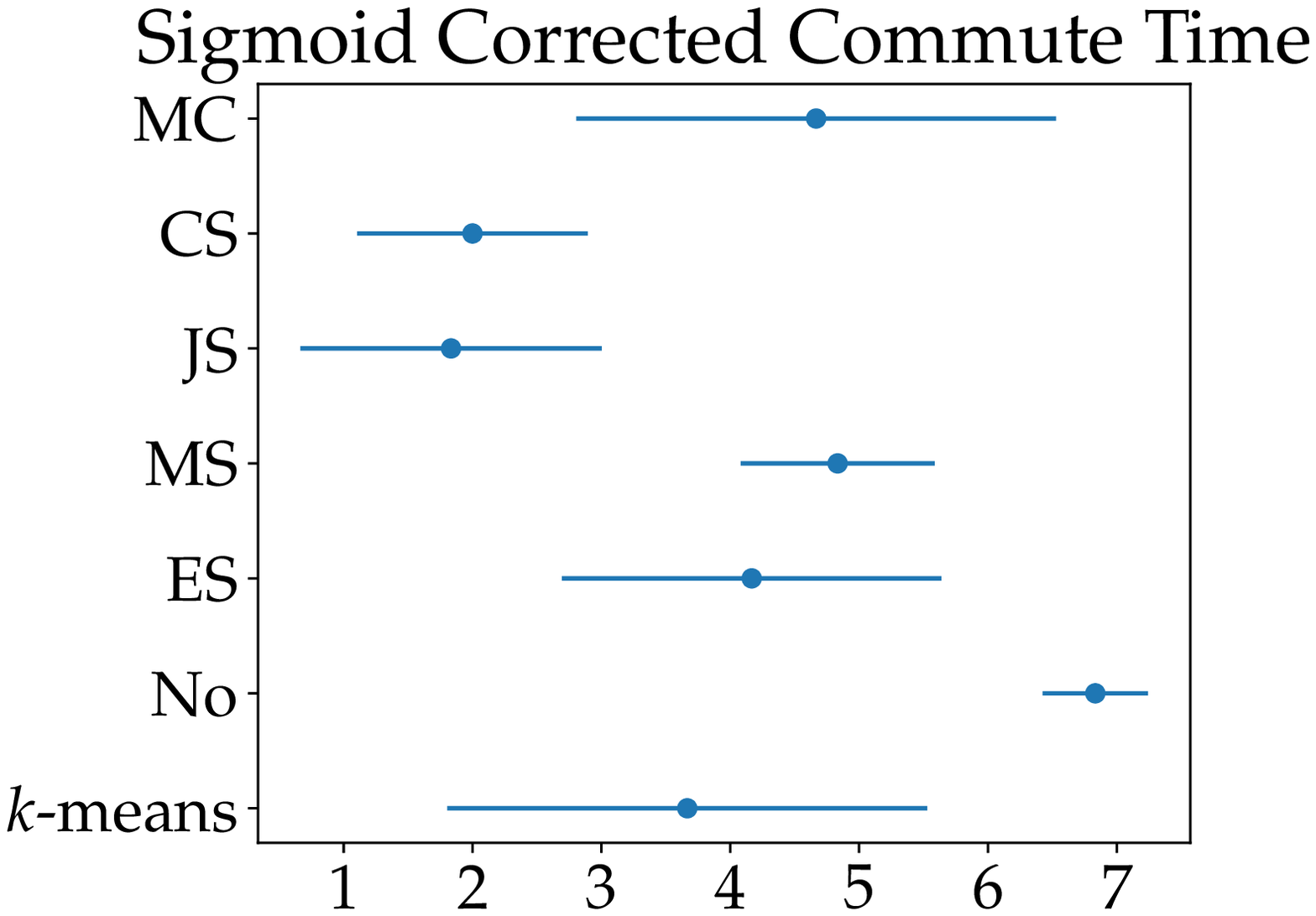}
\caption{Average rank and standard deviation for attribute similarity measures and $k$-means} 
\label{fig:sim-measures-means}
\end{figure}

Not all tested attribute similarity measures have shown good clustering quality. Figure \ref{fig:sim-measures-means} presents the average rank and standard deviation for attribute similarity measures and $k$-means. The rank is averaged over 6 datasets. The figure contains 5 graphs: one for each proximity measure. 

One can see that the Cosine Similarity and Extended Jaccard Similarity measures perform the best: they have the highest ranks for all the proximity measures. As for plain proximity measures, which are not combined with any similarity measure, they have one of the lowest ranks. The performance of the Matching Coefficient, the Manhattan Similarity, and the Euclidean Similarity measures varies for different proximity measures.

In Table \ref{table:best-pairs}, the top-performing combinations of a proximity measure and a similarity measure are presented. As can be seen, the undisputed leader is Free Energy combined with the Cosine Similarity measure.

\begin{table}
\caption{The top-performing pairs of proximity measure and similarity measure}
\label{table:best-pairs}
\centering
\begin{tabular}{p{0.5cm}p{3cm}p{3cm}p{2cm}}
\noalign{\smallskip}\hline\noalign{\smallskip}
№ & Proximity measure & Similarity measure & Average rank  \\
\noalign{\smallskip}\hline\noalign{\smallskip}
1 & FE & CS & 2.833 \\
2 & FE & JS & 6.333 \\
3 & Communicability & CS & 6.667 \\
4 & SCCT & JS & 7.333 \\
5 & SCCT & CS & 7.667 \\
6 & Communicability & JS & 8.333 \\
7 & PR & CS & 8.333 \\
8 & Heat & CS & 8.667  \\
\hline
\end{tabular}
\end{table}

\section{Conclusion}
In this paper, we investigated the possibility of applying proximity measures for community detection in attributed networks. We studied a number of proximity measures, including Communicability, Heat, PageRank, Free Energy, and Sigmoid Corrected Commute-Time. Attribute information was embedded into proximity measures using several attribute similarity measures, i.e., the Matching Coefficient, the Cosine Similarity, the Extended Jaccard Similarity, the Manhattan Similarity, and the Euclidean Similarity.

According to the results of the experiments, taking node attributes into account when measuring proximity improves the efficiency of proximity measures for community detection. Not all attribute similarity measures perform equally well. The top-performing attribute similarity measures were the Cosine Similarity and Extended Jaccard Similarity.

Future studies may address the problem of choosing the optimal $\beta$ in \eqref{eq:similarity-measure}. Another area for future research is to find more effective attribute similarity measures.

\bibliography{proximity_attributed}

\begin{thebibliography}{10}
\providecommand{\url}[1]{\texttt{#1}}
\providecommand{\urlprefix}{URL }

\bibitem{cheb-on-kernels}
Avrachenkov, K., Chebotarev, P., Rubanov, D.: Kernels on graphs as proximity
  measures. In: International Workshop on Algorithms and Models for the
  Web-Graph. LNCS, vol. 10519, pp. 27--41. Springer (2017)

\bibitem{transformations}
Aynulin, R.: Efficiency of transformations of proximity measures for graph
  clustering. In: International Workshop on Algorithms and Models for the
  Web-Graph. LNCS, vol. 11631, pp. 16--29. Springer (2019)

\bibitem{bothorel2015clustering}
Bothorel, C., Cruz, J.D., Magnani, M., Micenkova, B.: Clustering attributed
  graphs: Models, measures and methods. Network Science  3(3),  408--444 (2015)

\bibitem{cheb-forest-kernel}
Chebotarev, P.Y., Shamis, E.: On the proximity measure for graph vertices
  provided by the inverse {L}aplacian characteristic matrix. In: 5th Conference
  of the International Linear Algebra Society, Georgia State University,
  Atlanta. pp. 30--31 (1995)

\bibitem{cheb-walk}
Chebotarev, P.: The walk distances in graphs. Discrete Applied Mathematics
  160(10-11),  1484--1500 (2012)

\bibitem{chunaev2020community}
Chunaev, P.: Community detection in node-attributed social networks: a survey.
  Computer Science Review  37,  100286 (2020)

\bibitem{applications}
Costa, L.d.F., Oliveira~Jr, O.N., Travieso, G., Rodrigues, F.A., Villas~Boas,
  P.R., Antiqueira, L., Viana, M.P., Correa~Rocha, L.E.: Analyzing and modeling
  real-world phenomena with complex networks: A survey of applications.
  Advances in Physics  60(3),  329--412 (2011)

\bibitem{dang2012community}
Dang, T., Viennet, E.: Community detection based on structural and attribute
  similarities. In: International Conference on Digital Society (ICDS). pp.
  7--12 (2012)

\bibitem{DezaDeza16}
Deza, M.M., Deza, E.: Encyclopedia of Distances. Fourth Edition, Springer,
  Berlin (2016)

\bibitem{dijkstra1959note}
Dijkstra, E.W., et~al.: A note on two problems in connexion with graphs.
  Numerische Mathematik  1(1),  269--271 (1959)

\bibitem{comm-distance}
Estrada, E.: The communicability distance in graphs. Linear Algebra and its
  Applications  436(11),  4317--4328 (2012)

\bibitem{fouss2012experimental}
Fouss, F., Francoisse, K., Yen, L., Pirotte, A., Saerens, M.: An experimental
  investigation of kernels on graphs for collaborative recommendation and
  semisupervised classification. Neural Networks  31,  53--72 (2012)

\bibitem{comm-distance-2}
Fouss, F., Yen, L., Pirotte, A., Saerens, M.: An experimental investigation of
  graph kernels on a collaborative recommendation task. In: Sixth International
  Conference on Data Mining (ICDM'06). pp. 863--868. IEEE (2006)

\bibitem{girvan2002community}
Girvan, M., Newman, M.E.: Community structure in social and biological
  networks. Proceedings of the National Academy of Sciences  99(12),
  7821--7826 (2002)

\bibitem{ari-hubert}
Hubert, L., Arabie, P.: Comparing partitions. Journal of Classification  2(1),
  193--218 (1985)

\bibitem{comparison-logarithmic}
Ivashkin, V., Chebotarev, P.: Do logarithmic proximity measures outperform
  plain ones in graph clustering? In: International Conference on Network
  Analysis. PROMS, vol. 197, pp. 87--105. Springer (2016)

\bibitem{jain2010data}
Jain, A.K.: Data clustering: 50 years beyond $k$-means. Pattern Recognition
  Letters  31(8),  651--666 (2010)

\bibitem{kivimaki2014developments}
Kivim{\"a}ki, I., Shimbo, M., Saerens, M.: Developments in the theory of
  randomized shortest paths with a comparison of graph node distances. Physica
  A: Statistical Mechanics and its Applications  393,  600--616 (2014)

\bibitem{heat-kernel}
Kondor, R., Lafferty, J.: Diffusion kernels on graphs and other discrete input
  spaces. In: International Conference on Machine Learning. pp. 315--322 (2002)

\bibitem{lu2003link}
Lu, Q., Getoor, L.: Link-based classification. In: Proceedings of the 20th
  International Conference on Machine Learning (ICML-03). pp. 496--503 (2003)

\bibitem{spectral-tutorial}
von Luxburg, U.: A tutorial on spectral clustering. Statistics and Computing
  17(4),  395--416 (2007)

\bibitem{luxburg2010getting}
von Luxburg, U., Radl, A., Hein, M.: Getting lost in space: Large sample
  analysis of the resistance distance. In: Advances in Neural Information
  Processing Systems. pp. 2622--2630 (2010)

\bibitem{macqueen1967some}
MacQueen, J., et~al.: Some methods for classification and analysis of
  multivariate observations. In: Proceedings of the fifth Berkeley Symposium on
  Mathematical Statistics and Probability. pp. 281--297. Oakland, CA, USA
  (1967)

\bibitem{ari-best}
Milligan, G.W., Cooper, M.C.: A study of the comparability of external criteria
  for hierarchical cluster analysis. Multivariate Behavioral Research  21(4),
  441--458 (1986)

\bibitem{neville2003clustering}
Neville, J., Adler, M., Jensen, D.: Clustering relational data using attribute
  and link information. In: Proceedings of the Text Mining and Link Analysis
  Workshop, 18th International Joint Conference on Artificial Intelligence. pp.
  9--15 (2003)

\bibitem{pagerank}
Page, L., Brin, S., Motwani, R., Winograd, T.: The pagerank citation ranking:
  Bringing order to the web. Tech. rep., Stanford InfoLab (1999)

\bibitem{rand}
Rand, W.M.: Objective criteria for the evaluation of clustering methods.
  Journal of the American Statistical Association  66(336),  846--850 (1971)

\bibitem{ruan2013efficient}
Ruan, Y., Fuhry, D., Parthasarathy, S.: Efficient community detection in large
  networks using content and links. In: Proceedings of the 22nd International
  Conference on World Wide Web. pp. 1089--1098 (2013)

\bibitem{sen2008collective}
Sen, P., Namata, G., Bilgic, M., Getoor, L., Galligher, B., Eliassi-Rad, T.:
  Collective classification in network data. AI Magazine  29(3),  93--93 (2008)

\bibitem{sharpe1967solution}
Sharpe, G.: Solution of the (m+1)-terminal resistive network problem by means
  of metric geometry. In: Proceedings of the First Asilomar Conference on
  Circuits and Systems, Pacific Grove, CA. pp. 319--328 (1967)

\bibitem{shi2000normalized}
Shi, J., Malik, J.: Normalized cuts and image segmentation. IEEE Transactions
  on Pattern Analysis and Machine Intelligence  22(8),  888--905 (2000)

\bibitem{comparison-felix}
Sommer, F., Fouss, F., Saerens, M.: Comparison of graph node distances on
  clustering tasks. In: International Conference on Artificial Neural Networks.
  LNCS, vol. 9886, pp. 192--201. Springer (2016)

\bibitem{vsulc2014evaluation}
Sulc, Z., Řezanková, H.: Evaluation of recent similarity measures for
  categorical data. In: Proceedings of the 17th International Conference
  Applications of Mathematics and Statistics in Economics. Wydawnictwo
  Uniwersytetu Ekonomicznego we Wroc{\l}awiu, Wroclaw. pp. 249--258 (2014)

\bibitem{tan2016introduction}
Tan, P.N., Steinbach, M., Kumar, V.: Introduction to Data Mining. Pearson
  Education India (2016)

\bibitem{yang2013community}
Yang, J., McAuley, J., Leskovec, J.: Community detection in networks with node
  attributes. In: 2013 IEEE 13th International Conference on Data Mining. pp.
  1151--1156. IEEE (2013)

\bibitem{zhou2009graph}
Zhou, Y., Cheng, H., Yu, J.X.: Graph clustering based on structural/attribute
  similarities. Proceedings of the VLDB Endowment  2(1),  718--729 (2009)

\end{thebibliography}
\end{document}